\documentclass[10pt]{amsart}

\usepackage[margin=1in]{geometry}
\usepackage{amsmath,amssymb}
\usepackage{graphicx}
\usepackage[table]{xcolor}
\usepackage{booktabs}
\usepackage{array}
\usepackage{tikz}
\usetikzlibrary{positioning,arrows.meta}
\usepackage{caption}
\usepackage{float}
\usepackage{microtype}
\usepackage{cite}
\usepackage[colorlinks=true,allcolors=blue]{hyperref}

\graphicspath{{figures/}}

\title{Tensor analysis for lipid transport}

\author[Pellegrinelli]{Ilaria Pellegrinelli\textsuperscript{1,2,3,4}}
\author[Galgano]{Vincenzo Galgano\textsuperscript{1,2,3}}
\author[Lennartz]{Mathilda Lennartz\textsuperscript{1}}
\author[Nadler]{Andr\'{e} Nadler\textsuperscript{1,5}}
\author[Harrington]{Heather A. Harrington\textsuperscript{1,2,3,6}}

\address{
\textsuperscript{1} Max Planck Institute of Molecular Cell Biology and Genetics (MPI-CBG), Dresden, Germany\\
\textsuperscript{2} Center of Systems Biology Dresden (CSBD), Dresden, Germany\\
\textsuperscript{3} Faculty of Mathematics, Technische Universit\"at Dresden (TUD), Dresden, Germany\\
\textsuperscript{4} Faculty of Mathematics, University of Trento, Trento, Italy\\
\textsuperscript{5} Cluster of Excellence Physics of Life, TU Dresden, Dresden, Germany\\
\textsuperscript{6} Mathematical Institute, University of Oxford, Oxford, UK
}

\date{} 

\newcolumntype{+}{!{\vrule width 2pt}}
\newlength\savedwidth

\begin{document}

\begin{abstract}
High-dimensional biological datasets with molecular, spatial, and temporal dimensions are increasingly common. However, their analysis requires approaches that can integrate multiple data axes, accommodate noisy and missing measurements, and capture both dynamic interactions and localization changes. To address this, we provide an end-to-end tensor analysis pipeline that handles sparsity by employing tensor decomposition methods (HOSVD and CP) that are augmented with a binary mask for missing data and a framework for measurement error. We showcase this on a three-dimensional mammalian lipid transport dataset depending on lipid identities, organelle localizations, and time-series abundances. Our approach successfully identifies specific lipid-organelle pairs undergoing rapid temporal evolution, uncovers modules of lipids that co-vary across organelles and time, and extracts latent factors representing global redistribution trajectories. Direct comparison with a previous kinetic ODE model confirms that the tensor decompositions faithfully reproduce key lipid flux features.
\end{abstract}

\maketitle

\section{Introduction}

Modern biological research increasingly relies on multi-omic techniques and high-throuput microscopy, resulting in high-dimensional datasets that capture observations of molecular details of living matter at unprecendened scale, but often exhibit higher noise levels and missing values. The spatio-temporal behaviour of such underlying biological systems and networks is driven by complex multi-dimensional interactions between variables. Analyses of such datasets typically either focus on dynamic species interactions (biochemical reactions and binding events) or the spatial localisation of biomolecules. Combining such information into a coherent data analysis framework (e.g. changing biomolecule localizations over time during transport processes) remains a challenge.

Encoding data in a multi-indexed array (eg Rubik's cube) or a tensor is a natural mathematical tool to store information related to biological interactions between multiple variables. The mathematical data analysis of tensors differs from well-known PCA or other decomposition methods established for matrices of data. Over the past decade, significant progress has improved the machinery for decomposing and interpreting tensors, both in theory and computations \cite{kolda2009tensor, ponnapalli2011higher, rabanser2017introduction} as well as biological applications \cite{seigal2019tensor,ahern2022blood, alter2004integrative, tan2024structure,Khoshnam2025TensorOmics}. The two most popular tensor decomposition methods are higher order singular value decomposition (HOSVD) \cite{de2000multilinear} and canonical polyadic (CP) decomposition \cite{carroll1970analysis, harshman1970foundations}. \\
Here, we provide an end-to-end tensor data analysis pipeline that handles sparse and noisy data, and then employs decompositions to provide interpretable results. We handle the missing values using a binary multi-dimensional mask, and we perform a diagnostic on the convergence. We include ideas that take into consideration the measurement error attached to the values of the tensor. After performing a tensor decomposition, the remaining challenge is to extract information that is biologically meaningful from the components. We provide a qualitative and quantitative procedure to interpret the outputs from a biological perspective. 

We showcase our pipeline by considering a three-dimensional dataset containing the mean abundances of membrane lipids across cellular organelles at different times. 
Membrane lipids are fundamental to cellular biochemistry \cite{membranelipid}. They are the primary means of the cell to compartmentalize its biochemistry, as they are the structural building blocks required for the formation of organelle membranes. Each organelle membrane has its own dedicated lipid composition that has to be carefully maintained. Mammalian cells contain thousands of molecularly distinct lipid species that have to be sorted into organelle membranes by cellular transport processes \cite{harayama2018understanding}. The cellular lipid flux network that describes the movement of lipid building blocks through chemical (metabolism) and physical space (transport between organelle membranes) comprises many hundreds of proteins, and this complexity is the main reason that the molecular mechanisms of lipid transport are still not well understood \cite{kim2023lipid}.
Using chemical probes, we recently developed a quantitative lipid imaging approach that enables the acquisition of lipid transport time series data in mammalian cells (Iglesias-Artola et al, \cite{iglesias}). 

This yielded a dataset comprising lipid chemical identities, time and lipid abundances in organelles: in particular, the multidimensionality of the dataset arises from the times and locations associated to each lipid species. By fitting a mass action kinetic model encompassing the major lipid transport routes to these data, we quantified inter-organelle transport rates and found that non-vesicular transport modes convey most of the selectivity. However, this approach cannot be scaled easily and additional data analysis strategies are needed to tackle increasingly complex datasets. 

Starting from the aforementioned dataset, we show how to apply the desired tensor decompositions and how to extract meaningful biological considerations from the outputs. All procedures are scalable and can be generalized to any multi-dimensional dataset. 
By exploiting tensor decomposition methods we focus on three biological questions: \begin{enumerate}
    \item[Q1] Which lipids co-vary across organelles and over time? 
    \item[Q2] Which (lipid, organelle) pairs correspond to the fastest concentration changes over time? 
    \item[Q3] What are the interpretable latent factors that summarize global and/or specific patterns, such as trajectories of lipid redistribution?
\end{enumerate}

Throughout our analysis, we assess whether HOSVD and CP decomposition can capture relevant features of lipid transport from three-dimensional biological data. We re-analyze a subset of the lipid flux dataset of Iglesias-Artola et al. \cite{iglesias}, previously analyzed using a kinetic ODE model, applying tensor decompositions. The direct comparison of both approaches shows that tensor factorizations faithfully recover key features of lipid flux and thus can be used as a first step to inform the choice of more precise follow-up kinetic analysis of network substructures. Our results establish tensor decomposition as a valid and scalable method for analysing complex lipid flux datasets.

\section*{Materials and methods}

In this section, we describe the end-to-end pipeline consisting of two tensor decompositions that we apply, and we explain how to construct the tensor starting from the raw biological data. We recall that tensors are multi-indexed arrays. The number of indices, or parameters or \textit{modalities}, is also called the \textit{order} of the tensors. For instance, matrices are tensors of order two, while Rubik's cubes are tensors of order three. A tensor of order $d$ is a natural mathematical structure to encode information about a $d$-dimensional dataset. Every tensor can be decomposed as sum of \textquotedblleft simpler\textquotedblright \space (i.e. of lower complexity) tensors, which are called rank-one tensors. Finding a decomposition of minimal length is informative about the total complexity of the dataset, and allows one to isolate the most meaningful information of the dataset. In the case of matrices, an exact decomposition into rank-one matrices is guaranteed by the singular value decomposition, also known as principal component analysis (PCA). However, for higher-order tensors there is no direct analogue. Two algorithms that decompose, or at least approximate, tensors are the following. 

\subsection*{HOSVD}

The Higher Order Singular Value Decomposition builds on the matrix Singular Value Decomposition in order to decompose higher-order tensors. It is defined iteratively as follows: after fixing a modality of the tensor, it flattens the tensors into a matrix and performs a Singular Value Decomposition of such matrix. As a result, each modality yields to a factor matrix whose columns are the principal directions of variation of the fixed modality.

\subsection*{CP decomposition}
The CP decomposition approximates a tensor as a sum of rank-one tensors for a chosen number of summands. Each one of them is called factor and represents a multi-dimensional direction. 
The output of the CP decomposition looks similar to the classic PCA, although the resulting factors are in general not orthogonal and not uncorrelated. In CP decomposition, choosing the optimal number of factors is a critical challenge, which governs the trade-off between the discovery of macroscopic patterns and the extraction of highly specific behaviors of sub-clusters.

\subsection*{Consensus analysis for rank selection}
To determine the best number of factors for the CP decomposition, we adopt an empirical procedure. More precisely, we follow the consensus-based approach described in \cite{chafamo2023robust}, which evaluates both the fitability and the stability of the biological factors obtained from the decomposition.

Two major problems of CP decomposition are the invariance up to permutation of the factors and the possible convergence to local minima. To overcome these obstacles, for each candidate rank $R$ we run the decomposition $15$ times using random initializations. We concatenate the factor matrices across all the runs and normalize them. Next, to achieve a global minima, we first apply the Local Outlier Factor algorithm to remove outliers. Then, to remove the dependency on permutations of the factors of the decomposition across different runs, we cluster the remaining columns using the k-means algorithm, with $k=R$: this step aligns the components across different runs, solving the permutation ambiguity. 
Finally, we extract the medians from each cluster and use them as initial values for a definitive final run of the decomposition. 

The optimal number of factors is selected according to three metrics:
\begin{itemize}
\item \textbf{Explained Variance:} measures the quality of the approximation. We look for the ``elbow" in the variance curve.
\item \textbf{Cophenetic Correlation:} measures the stability and reproducibility of the latent biological signals across the different runs. If a pattern corresponds to a true biological signal, then it is likely to appear in multiple runs.
\item \textbf{Silhouette Coefficient:} measures the separation of the factor clustering.
\end{itemize}

\begin{figure}[H]
    \begin{center}
    \includegraphics[width=.65\textwidth, height=.285\textheight]{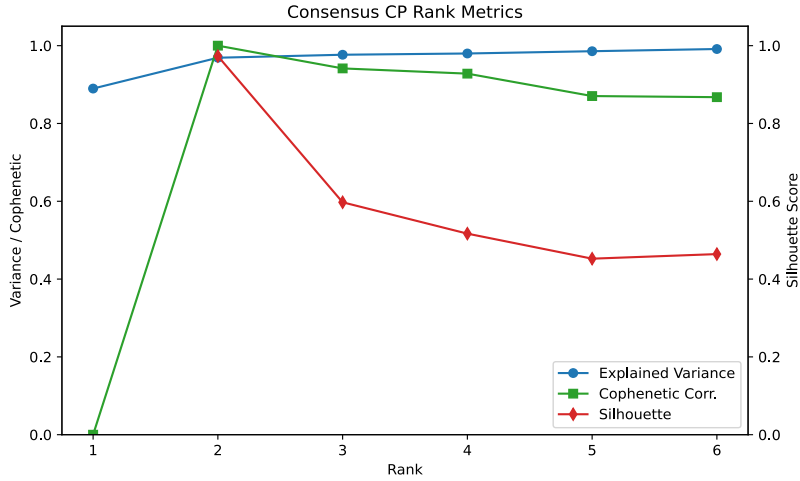}
    \caption{The values of the three metrics of the Consensus approach, for different ranks.}
    \label{fig: fig_consensus}
    \end{center}
\end{figure}

\noindent According to the plot shown in Figure \ref{fig: fig_consensus}, the optimal number of factors is $R=2$. However, to obtain a finer granularity and precision, we select $R=3$, which still maintains high values for each one of the three metrics. 
For completeness, in the Supporting Information section we include the outputs of CP decomposition for ranks $2$ (Figure \ref{fig: cp_2}) and $4$ (Figure \ref{fig: cp_4}).

\subsection*{Empirical confidence interval for HOSVD}
Given the measurement error available, we can handle noisy data in the tensor analysis. We propose a bootstrapping procedure which improves HOSVD and enables us to assess robustness and sensitivity. We sample each entry of the tensor from a Gaussian distribution, with mean given by the mean concentration of the lipid species and standard deviation equal to the measurement error. We repeat this construction of the tensor $200$ times, hence obtaining multiple realizations of the tensor; for each one, we perform HOSVD and compute the temporal heatmap as described later in the Results. This serves as an empirical confidence interval of how different the outputs of the heatmap can be: if a measurement is unreliable, then the range of possible values is wider.

\subsection*{Binary mask for CP decomposition} To handle missing data, we associate a multi-dimensional binary mask to the tensor, which enables one to iterate the minimization step of the loss function only on the true observed values. This represents
an advantage with respect to the standard methods: it preserves the inherent multi-dimensionality
of the data and also avoids the introduction of noise or biases by sidestepping imputation of the
missing entries. 

\section*{Biological description of data}
Recently, near-native bifunctional (crosslinkable and clickable) lipid probes allow one to trace lipid localizations at the nanoscale.
Using such bifunctional probes, a subset of authors recently developed a quantitative lipid imaging approach that enables the measurement of lipid transport kinetics in mammalian cells \cite{iglesias}. Lipid localizations were tracked across three experimental dimensions, lipid species, organelle localization and time, yielding a three-dimensional dataset. By fitting a mass action kinetic model encompassing the major lipid transport routes to these data we quantified inter-organelle transport rates and found that non-vesicular transport modes convey most of the selectivity. Here we study the same lipid dataset with a tensor, complementing mass action analysis.

\section*{Results}
We summarise the pipeline to answer the three biological questions by first describing how to transform a dataset of spatiotemporal lipid species into a tensor, and then how to analyse it with tensor decomposition methods. We apply a HOSVD pipeline to answer the biological questions of detecting the species that spatio-temporally co-vary (Q1) , and of identifying which lipid-organelle pairs change concentration fastest over time (Q2). Moreover, using CP decomposition we address the problem of finding factors for lipid redistribution trajectories (Q3).

\subsection*{From spatiotemporal lipid species to tensor construction and pipeline}
We consider a dataset containing the mean abundances of lipids across cellular organelles at different times. This type of data is three-dimensional and hence admits a natural interpretation as a tensor of order three. The three modalities are: lipids, organelles, times. The value of the $(i,j,k)$-th entry represents the mean concentration of lipid $i$ in organelle $j$ at time $k$. The eight lipids are the phosphatidylcholines PC(Y16/16:0), PC(Y16:18:1), PC(16:0/Y16), PC(18:1/Y16), PC(Y16/20:4), PC(20:4/Y16), the sphingomyelin SM(d18:1/Y16) and the phosphatidic acid PA(18:1/Y16). The five organelles are the endoplasmic reticulum (ER), the endosome (Endo), the Golgi organelles, the mitochondria (Mito) and the plasma membrane (PM). The five measurement times are $t=4,30,60,120,240$ minutes.

\begin{figure}[ht]
    
\centering
\begin{tikzpicture}[
    font=\small,
    mainbox/.style={
        draw,
        rounded corners,
        align=center,
        minimum height=10mm,
        inner xsep=1mm,
        inner ysep=1mm
    },
    flowarrow/.style={
        -{Latex},
        thick,
        shorten <=1mm,
        shorten >=1mm
    }
]

\node[mainbox] (box1)
{Rank Selection\\[-1pt]
{(Consensus Analysis)}};

\node[mainbox, right=8mm of box1] (box2)
{Tensor Decomposition\\[-1pt]
{(CP or HOSVD)}};

\node[mainbox, right=8mm of box2] (box3)
{Noisy Data\\[-1pt]
{(Empirical Confidence)}};

\node[mainbox, right=8mm of box3] (box4)
{Missing Values\\[-1pt]
{(Binary Mask)}};

\draw[flowarrow] (box1.east) -- (box2.west);
\draw[flowarrow] (box2.east) -- (box3.west);
\draw[flowarrow] (box3.east) -- (box4.west);

\end{tikzpicture}
\vspace{1ex}
\caption{Tensor data analysis pipeline.}
    \label{fig:tensorpipeline}
\end{figure}
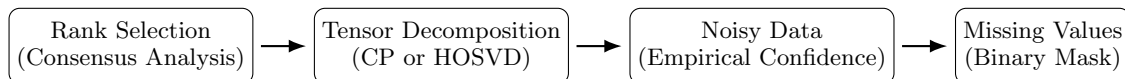

To deduce meaningful biological considerations from this tensor and handle problems that may arise when dealing with multi-dimensional biological data, we present an end-to-end pipeline from data to tensor analysis (employing two strategies, HOSVD and CP decomposition) to interpretation. The end-to-end tensor data analysis pipeline includes selecting number of components, tensor decomposition, handling noisy data and missing data, as summarized in Figure~\ref{fig:tensorpipeline}.

Two major data analysis obstacles in large biological datasets are measurement errors/variance and missing values.
To overcome the presence of \textit{noise}, we propose an empirical confidence interval for the heatmaps resulting from HOSVD, as a result of a bootstrapping procedure where we sample within the range of the mean concentration perturbed by the standard error of measurement for that entry. 
Concerning the problem of \textit{missing values}, we show that CP decomposition can be equivalently carried out by associating the tensor with a multi-dimensional binary mask, that allows to iterate the minimization step of the loss function only on the true observed values. This represents an advantage with respect to the standard methods: it preserves the inherent multi-dimensionality of the data and also avoids the introduction of noise or biases by sidestepping imputation of the missing entries. Moreover, the outputs of the decomposition can be successfully exploited to make predictions about the unknown missing entries.

\subsection*{(Q1) HOSVD identifies spatiotemporal dynamic lipid clusters} 
\begin{figure}[H]
    \centering
    \includegraphics[width=.6\textwidth, height=.285\textheight]{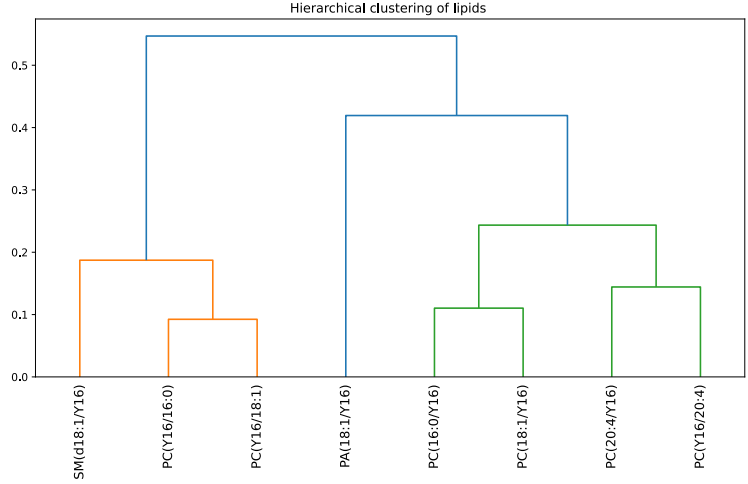}
    \caption{Lipid clusters from truncated HOSVD.}
    \label{fig: cluster_hosvd_trunc}
\end{figure}

We start by performing truncated HOSVD on the tensor.  As outputs, we obtain three factor matrices (one for each modality) containing the principal directions of variation and a smaller core tensor, encoding the reciprocal multi-dimensional interactions between the different variables.
We use these factors to divide the lipids into clusters (Figure \ref{fig: cluster_hosvd_trunc}), based on their normalized behavior in space and time (Q1).

The clusters of lipids can be interpreted in the following way. 
Lipids in the same cluster share correlated dynamics across organelles and time: they co-vary together.
Reassuringly, we find that structurally closely related species, in particular the phosphatidylcholines pairs PC(Y16/16:0) and PC(Y16:18:1), PC(16:0/Y16) and PC(18:1/Y16) and PC(Y16/20:4) and PC(20:4/Y16) cluster together, whereas the structurally dissimilar sphingomyelin SM(d18:1/Y16) and phosphatidic acid PA(18:1/Y16) lipids are more separated from other probes, reproducing similarities observed in the previous kinetic analysis. This indicates that even a global analysis of lipid dynamics can point to interesting biology.

\subsection*{(Q2) HOSVD determines lipids vary most in the PM and ER} 
\begin{figure}[H]
    \centering
   \includegraphics[width=1\textwidth, height=.285\textheight]{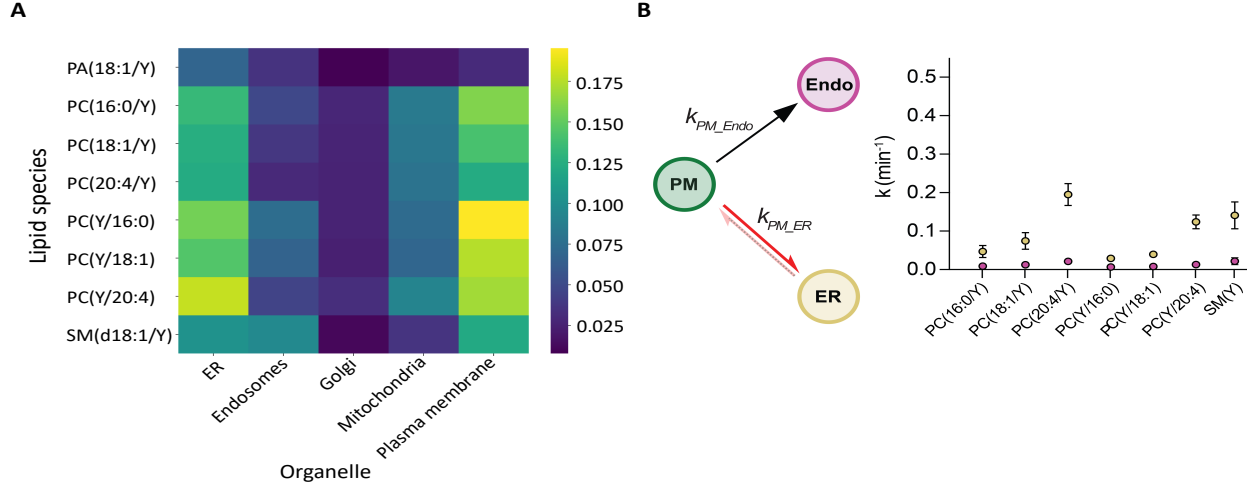}
   \caption{A) Temporal heatmap from truncated HOSVD. B) Left panel: Scheme of possible lipid transport pathways from the plasma membrane to either endosomes or the ER. Right panel: Rate constants of PM-ER transport are significantly higher than rate constants of PM-Endosome transport (reproduced from Iglesias-Artola et al \cite{iglesias}).}
   \label{fig: heatmap_hosvd_trunc}
\end{figure}

Using HOSVD, we produce a heatmap representing the temporal evolution of each (lipid, organelle)-pair (Figure \ref{fig: heatmap_hosvd_trunc}).
Mathematically, the temporal heatmap is the sum of the temporal factor matrices weighted by the entries of the core tensor, and projected on the (lipid, organelle) plane.
For completeness, we include the corresponding plots from full HOSVD in the Supporting Information section (Figures 
\ref{fig: heatmap_hosvd}). 
We also include HOSVD analysis with measurement errors (see Methods) to determine the empirical confidence interval and assess the results (see Figure \ref{fig: bootstrap}).

If a (lipid, organelle) pair has high contribution in the heatmap, it means that the concentration of the lipid in the organelle presents high fluctuations across time. 
Rows with similar global color distributions correspond to lipids that share similar temporal dynamics across organelles.
For all lipids except PA(18:1/Y16), the major changes across time are seen in the plasma membrane (PM) and the endoplasmic reticulum (ER). In particular, the most variable pairs are (PC(Y16/20:4), ER), (PC(Y16/16:0), PM), (PC(Y16/18:1), PM), (PC(Y16/20:4), PM). 
This is in line with our previous finding that the major retrograde lipid transport pathway is non-vesicular transport of lipids from the plasma membrane to the ER. Specifically, we previously found that for all probes the transport rates between the PM and the ER are the highest,
(Figure 3B), whereas the other possible route from the plasma membrane to endosomes is a minor contribution to overall lipid flux.
The lack of temporal signatures for PA(18:1/Y16) is fully in line with the experimental observations, as the redistribution of this particular lipid species is too fast to be captured with the experimental setup, it has reached its steady state distribution at the earliest time point.

To resume, from the factors of the HOSVD we deduce that the two organelles in which the most changes occur are PM and ER, and in particular the most variable pairs are (PC(Y16/20:4), ER), (PC(Y16/16:0), PM), (PC(Y16/18:1), PM), (PC(Y16/20:4), PM).

\subsection*{(Q3) Latent factors encode multidimensional biological patterns}
We use CP decomposition and interpret the factors to identify different spatial-temporal regimes underlying the dataset (Q3).
We start from a tensor and decompose it as a sum of factors,  representing the tensor products of a vector of each modality.
Each factor encodes a multi-dimensional biological pattern.
To align the mathematical analysis with the biological interpretation, we perform a non-negative CP decomposition: we enforce non-negativity of the factors, as the data represents concentrations of lipids, for which negative values would be biologically meaningless. 

Although visually similar to PCA, this decomposition presents fundamental differences, that makes it more tailored for the analysis of complex relationships between multiple variables. More specifically, we are not projecting or flattening, hence we do not lose the inherent multi-dimensional structure of data;
and the factors are, in general, correlated and not orthogonal, accurately portraying the typical correlated patterns that characterize true biological signals.

First, we perform the CP decomposition with 2 factors, which is the optimal number deduced from the previously described consensus-based approach.
Then we also try with, respectively, 3 and 4 factors. 
This will likely break global biological patterns, but at the same time this finer granularity will allow us to identify more specific patterns that only involve some lipids or organelles and that would remain hidden otherwise.

We look at the factors in the following way. 
Each factor is a tensor product of a lipid vector, an organelle vector and a time vector. 
An entry of high magnitude means that that element strongly contributes to that biological pattern; an entry of small magnitude can be ignored, since the attached signal is irrelevant. 
This procedure is justified by the fact that, under mild assumptions \cite{ten2002uniqueness}, CP decomposition is unique up to permutation and scaling of the factors, hence directly interpretable.

To get a better intuition, we look at the histograms of the entries of the three normalized factors of the CP decomposition, and explain how to use them to infer biological deductions, answering question (Q3).
The first factor captures and isolates plasma membrane pattern. At the initial time point, all lipids have a preferred localization at the plasma membrane, which reflects the experimental set-up, the lipid probes are delivered to the plasma membrane and subsequently re-distributed by the cellular lipid transport machinery. Again, the absence of this pattern for PA(18:1/Y16) is capturing the extremely fast transport of this particular lipid. The second and third factors capture the lipid steady state distributions with much higher lipid content in the endoplasmic reticulum (ER) and the mitochondria. Intriguingly, temporal differences are detected, with early occurrence for PC(Y16/20:4), PC(20:4/Y) and PA(18:1/Y16), which are in fact transported much faster compared to the other PC species analyzed here.

\begin{figure}[H]
   \centering \includegraphics[width=1\textwidth]{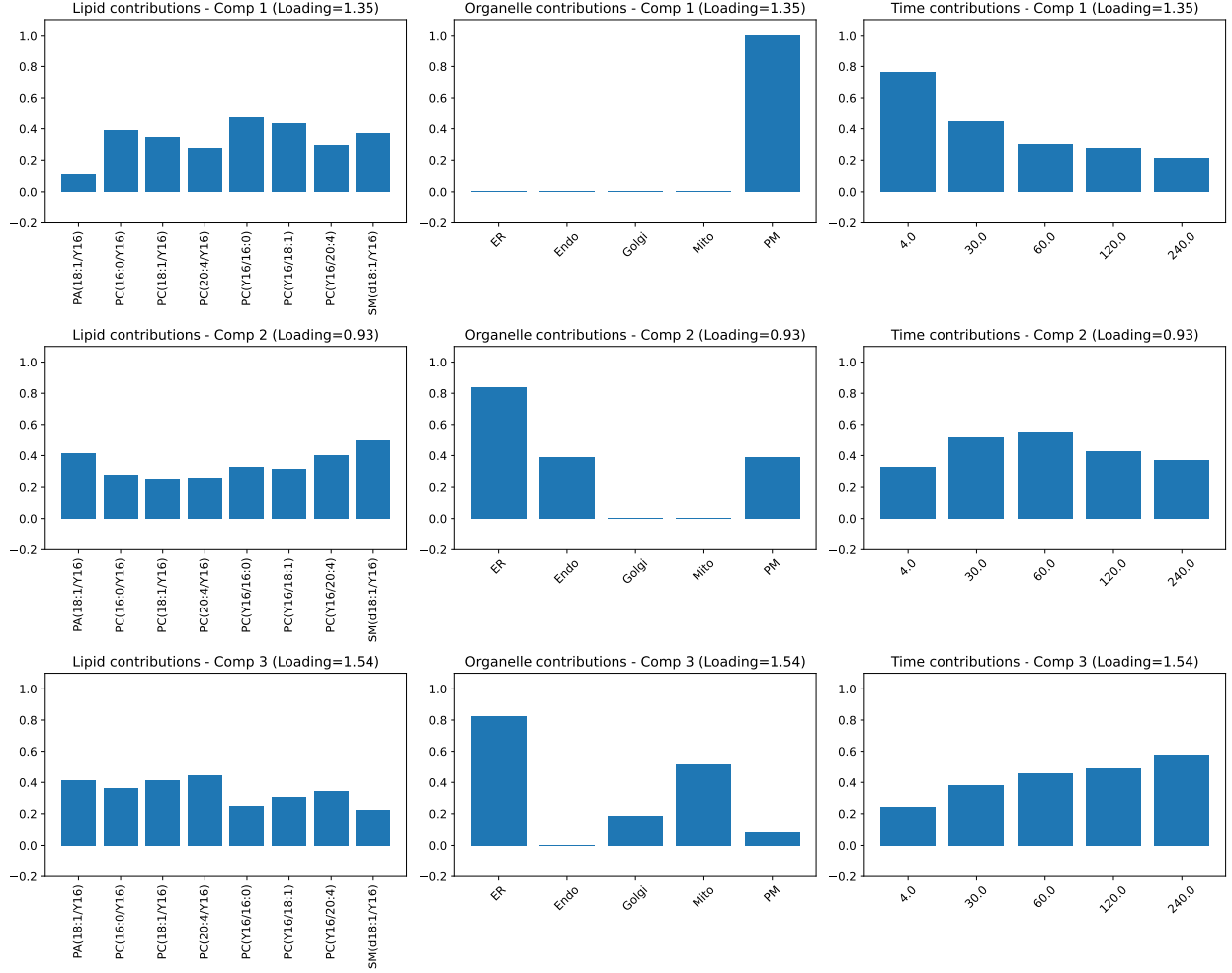}
   \caption{Histograms of the entries of the factors of CP decomposition with rank $3$.}
   \label{fig: cp_3_fact_1}
\end{figure}

\subsection*{Binary mask added to the tensor to handle missing values}

A good feature of tensor decompositions is their ability to successfully handle missing values. 
The idea is to couple the tensor with a binary mask, whose entries have value $1$ if the correspondent entry in the tensor is an observed value, or $0$ if the correspondent entry is missing. 
The optimization process (ALS with squared loss function) optimizes only with respect to the true observed values, with two structural advantages with respect to standard matrix methods. Namely, no loss of information as we do not need to remove rows of data for which some measurements are missing and no bias and noise introduction as we do not need to impute the missing values. 

\begin{figure}[H]
    \centering
    \includegraphics[width=0.3\linewidth]{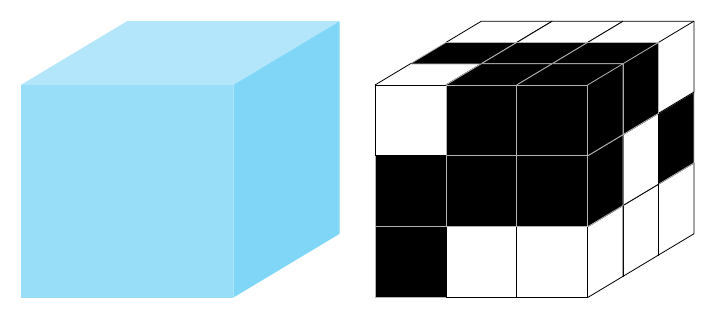}
    \caption{Binary mask associated with a tensor, to distinguish true and missing values.}
\end{figure}

We give a concrete example of the validity and robustness of this procedure, by artificially removing values from the tensor and performing again the CP decomposition. In this way, we are able to compare the outputs of the masked decomposition with the ground truth and quantify the goodness-of-fit and the accuracy of the predicted entries.
We remove $14$ randomly selected values in the tensor (thus $14/225 = 6.22 \%$ of values are missing). They are: lipid PC(Y16/16:0) in organelle ER at time 4; lipid PC(Y16/16:0) in organelle PM at time 60; lipid PC(16:0/Y16) in organelle ER at time 4; lipid PC(16:0/Y16) in organelle PM at time 60; lipid PA(18:1/Y16) in organelle Mito at time 30; lipid PC(Y16/18:1) in organelle Golgi at time 240; lipid PC(20:4/Y16) in organelle PM at time 4; lipid PC(20:4/Y16) in organelle Golgi at time 30; lipid PC(Y16/20:4) in organelle ER at time 60.
We repeat CP decomposition for ranks $2,3,4$ and in all settings we recover the biological patterns of the full complete case, qualitatively indicating a good performance of the method.

Here we present the results for rank $3$, matching the previous choice of the rank for CP decomposition on the full tensor. These factors recover the same patterns found in the previous analysis. Moreover, they can be used as a prediction for the missing parts. 
The relative reconstruction error, computed on the reconstructed missing values, corresponds to $14.12 \%$ (relative Frobenious norm of the error between the reconstructed missing entries and the true entries of the original tensor).

\begin{figure}[H]
    \centering
    \includegraphics[width=1\linewidth]{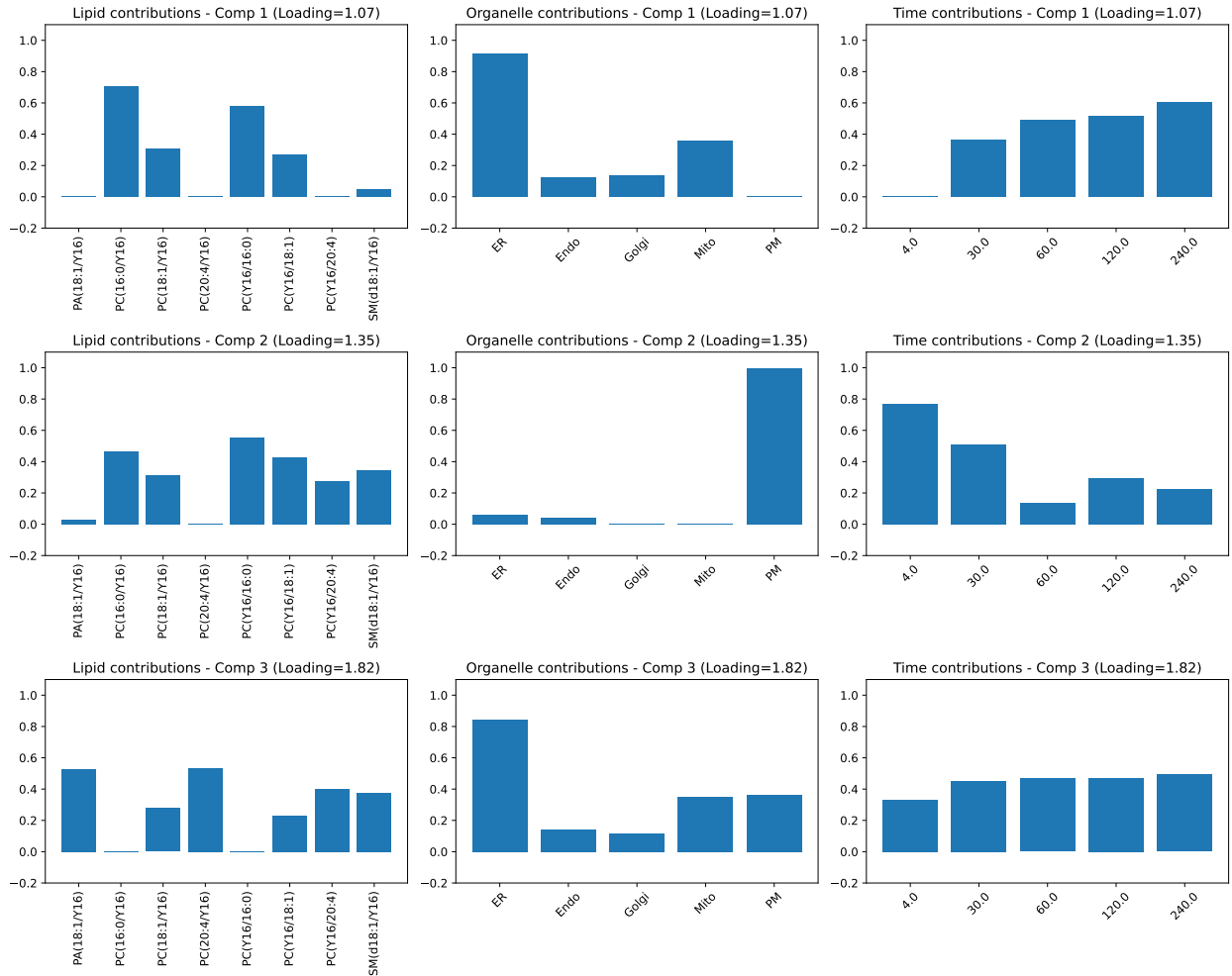}
     \caption{Histograms of the entries of the factors of CP decomposition with rank $3$ and missing values.}
\end{figure}

\section*{Discussion}

Classical hypothesis-driven kinetic models, such as ordinary differential equation (ODE)-based models, become increasingly difficult to apply as dimensionality grows. Each new experimental dimension requires explicit parameterization of additional interactions, which is often difficult to capture within one ODE model. 
Thus, for understanding cellular lipid biology on the molecular level, we require new analysis strategies that circumvent these problems.
We here test, whether tensor analysis can be used to capture key features of lipid transport dynamics in cells. Using two distinct tensor decompositions methodologies, we find that this approach correctly identifies key features of lipid transport (e.g. the prevalence of PM-ER lipid exchange and species-level differences in transport dynamics), and recapitulates central aspects of the experimental setup – e.g. the initial lipid localization of the lipid probes in the plasma membrane. It is noteworthy that this approach is independent of the nature of the dataset, and can be applied to de-noise, analyse and interpret other biological complex dynamical processes as well.

With the increasing availability of datasets with multiple types of molecules at finer spatio-temporal resolutions and whose interactions are dependent on spatial localization, additional mathematical data analysis methods are required. Consider nonlinear chemical reaction networks, which inherently encode assumptions about the biological system interactions; however, the well-mixed distribution of all reactants that underlies classical mass action kinetics lacks  transport and compartmentalization of large networks. Therefore future directions may include considering multi-indexed data structures, such as tensors presented here, which can incorporate such compartments, and fuse this data analysis with spatial chemical reaction network analysis.

\section*{Supporting information}

\begin{figure}[H]
    \centering
    \includegraphics[width=.8\textwidth, height=.285\textheight]{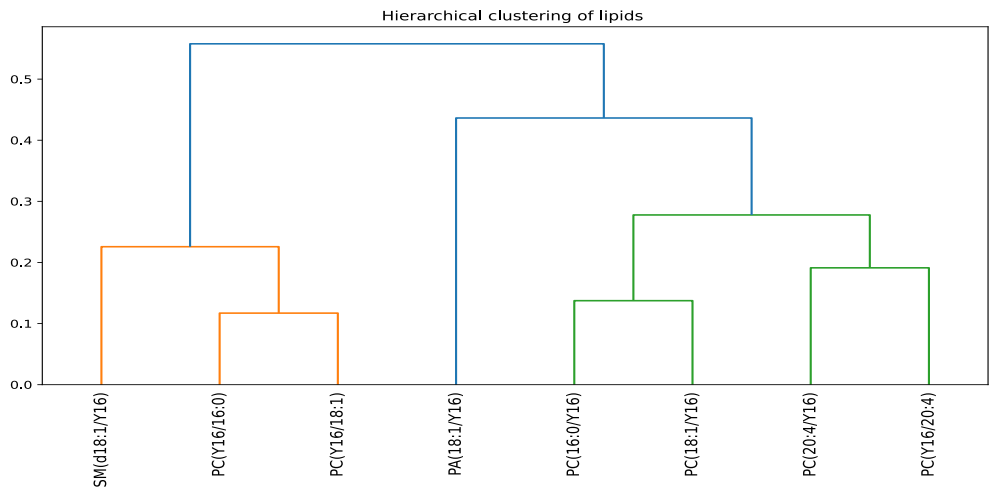}
    \caption{Lipid clusters from full HOSVD.}
    \label{fig:cluster_hosvd}
\end{figure}

\begin{figure}[H]
    \centering
   \includegraphics[width=.8\textwidth, height=.285\textheight]{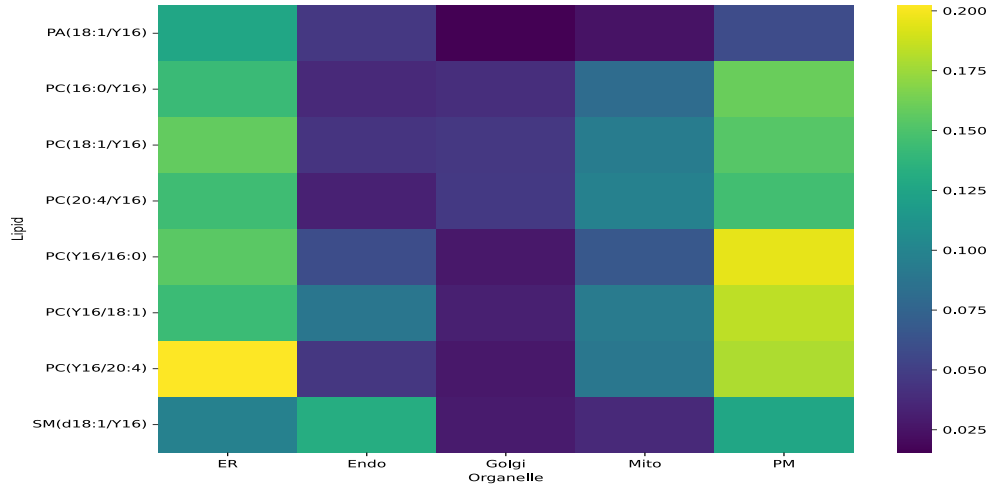}
   \caption{Temporal heatmap from full HOSVD.}
   \label{fig: heatmap_hosvd}
\end{figure}

\begin{figure}[H]
    \centering
    \includegraphics[width=1\textwidth, height=.285\textheight]{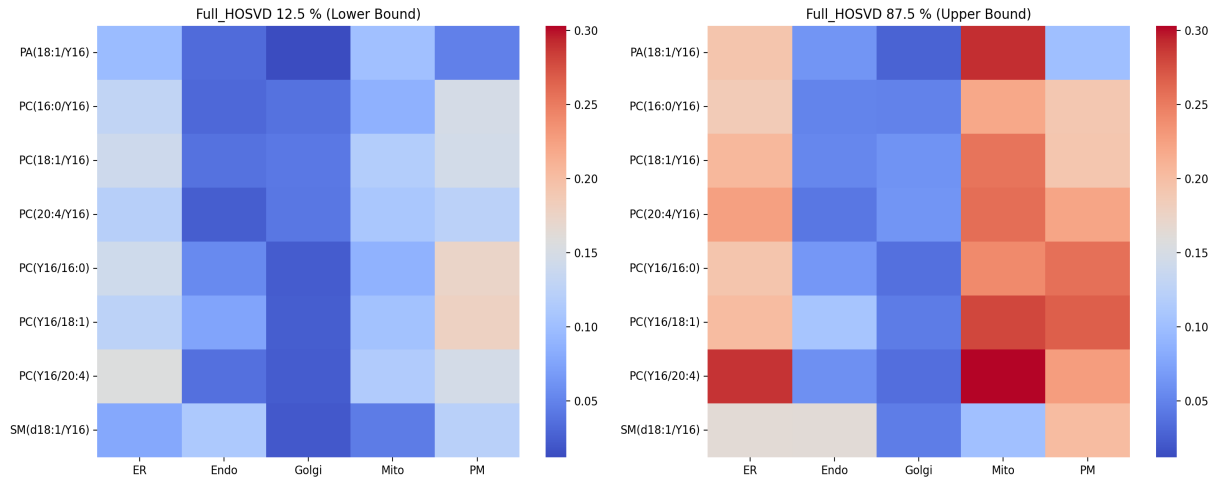}
    \caption{Empirical confidence bounds obtained from full HOSVD}
    \label{fig: bootstrap}
\end{figure}

\begin{figure}[H]
   \centering \includegraphics[width=1\textwidth]{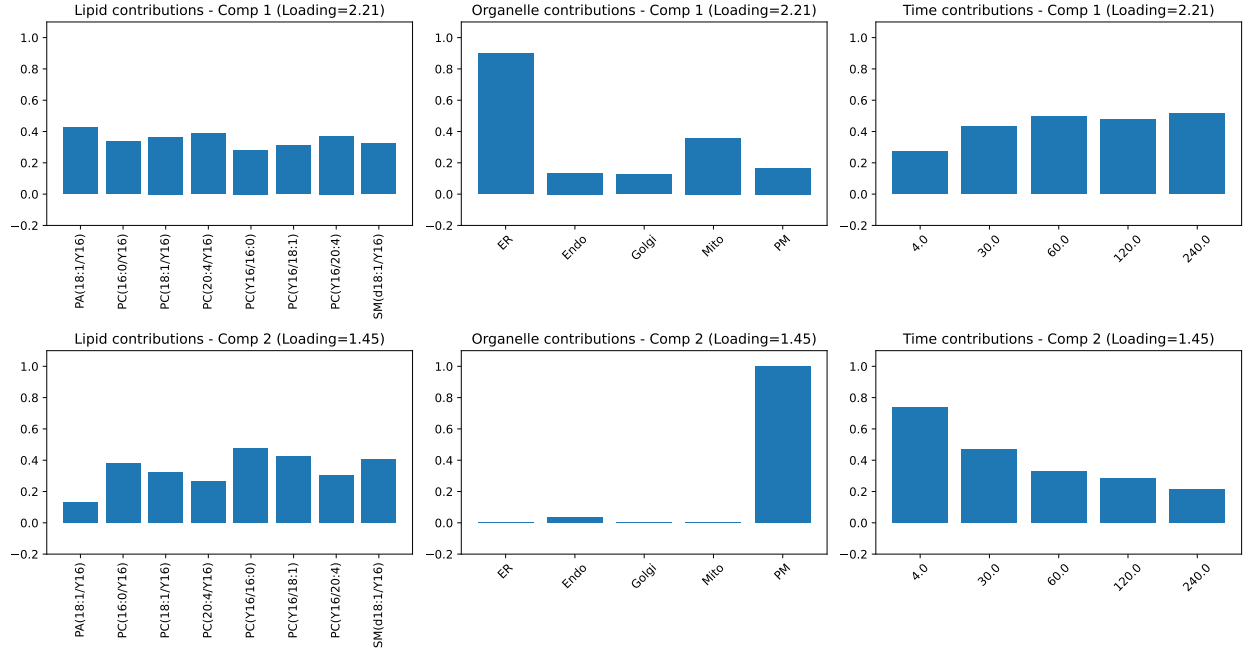}
   \caption{Histograms of the entries of the factors of rank-2 CP decomposition. The first factor represents a global pattern, shared among all lipids quite homogeneously. They tend to accumulate into ER (and Mito), homogeneously across time. The second factor captures a dynamic common to all lipids except for PA(18:1/Y16): in the initial times, they are accumulated in PM.}
   \label{fig: cp_2}
\end{figure}

\begin{figure}[H]
   \centering \includegraphics[width=0.9\textwidth]{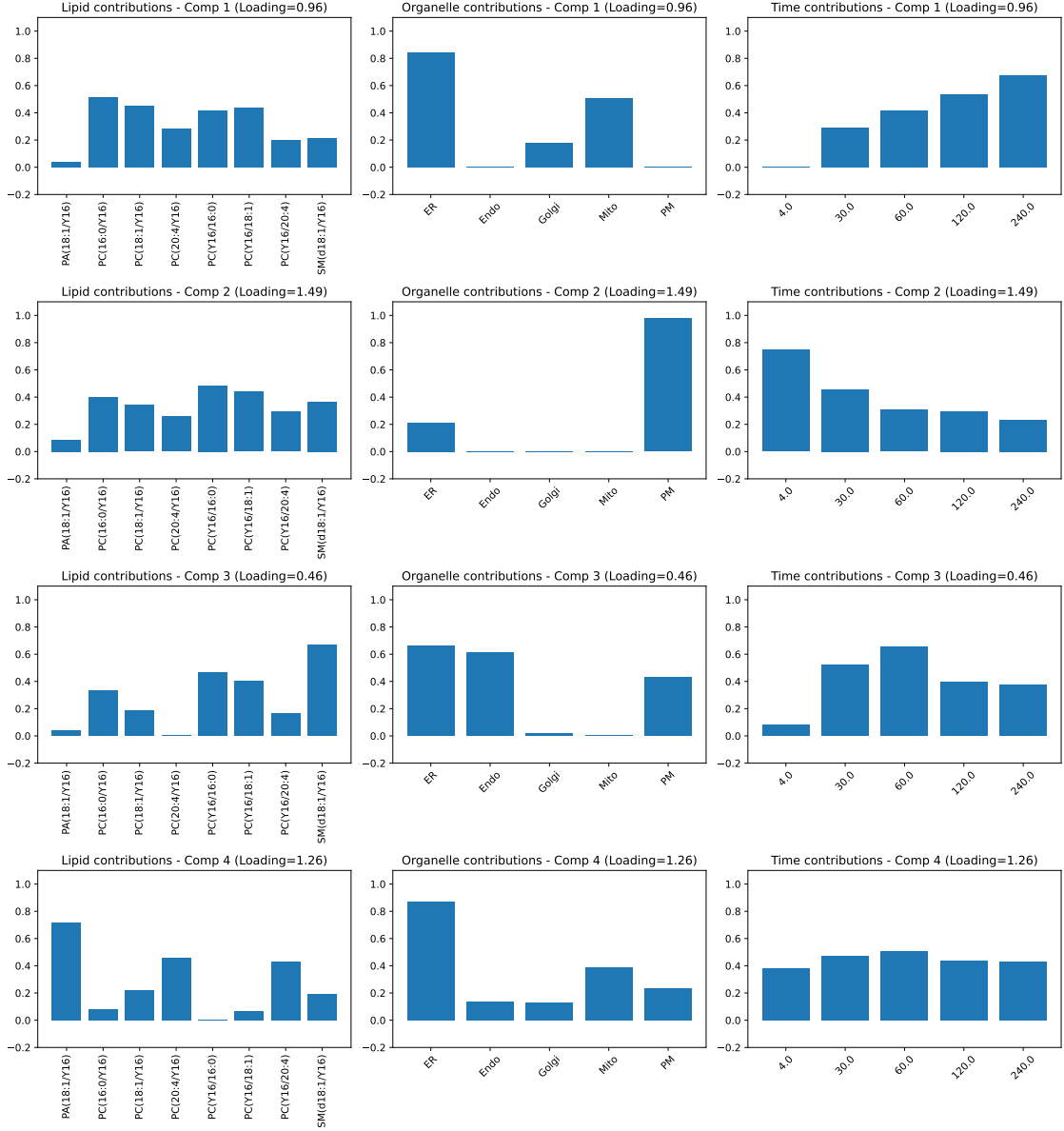}
    \caption{Histograms with the entries of the four factors of rank-$4$ CP decomposition.
    The first factor encodes an accumulation of PA(18:1/Y16) and (less strongly) PC(Y16/20:4), PC(20:4/Y16) into ER and Mito at initial times.
    The second factor is the usual global (except for PA(18:1/Y16)) dynamics of accumulation of PA(18:1/Y16) in PM at initial times. 
    The third factor seems to isolate a dynamical pattern involving mainly SM(d18:1/Y16), and also PC(Y16/16:0, PC(Y16/18:1)), that flows into ER, Endo and PM starting from time $30$.
    The fourth factor explains where the lipids migrate from PM: mainly into ER and Mito.}
   \label{fig: cp_4}
\end{figure}

We now provide the outputs of the analysis based on full HOSVD (Figures \ref{fig:cluster_hosvd}, \ref{fig: heatmap_hosvd}, \ref{fig: bootstrap}). 
The full and truncated approaches lead to equivalent biological interpretations. 
We have chosen to restrict our primary analysis to the truncated decomposition to mitigate noise and prevent the risk of overfitting.

Finally, we present the histograms of the entries of the factors of CP decomposition for ranks $2,4$.
We notice how they are coherent with the results obtained for rank $3$; however, they either merge two of the factors into one, providing a less precise view (rank $2$), or break biological signals into pieces (rank $4$). 

We also present the outputs of CP decomposition with missing values, handled via a binary tensor mask. Both for ranks $2$ and $4$, they match the results of the decomposition on the full tensor, successfully recovering the latent biological signals. 

\begin{figure}[H]
    \centering
    \includegraphics[width=0.8\linewidth]{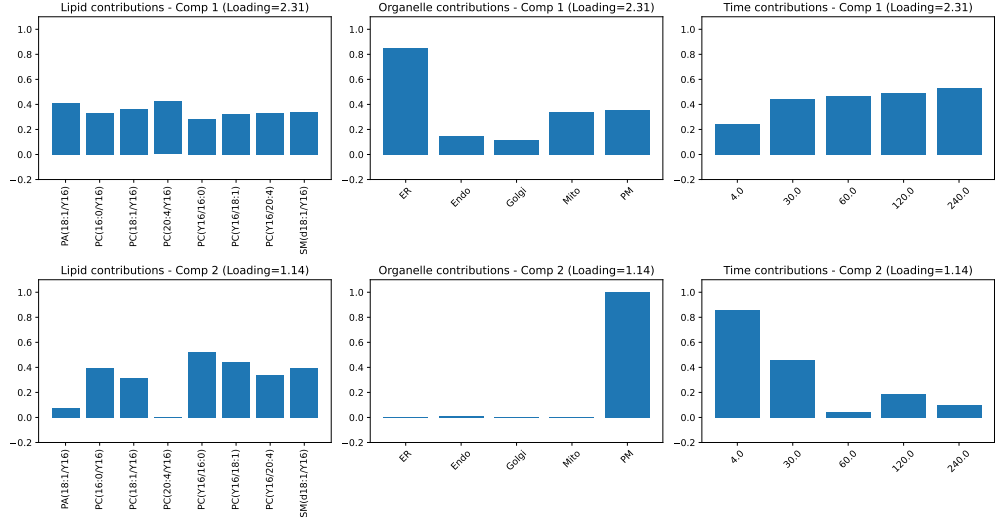}
     \caption{Histograms of the entries of the factors of rank-$2$ CP decomposition with missing values.}
\end{figure}

\begin{figure}[H]
    \centering
    \includegraphics[width=0.8\linewidth]{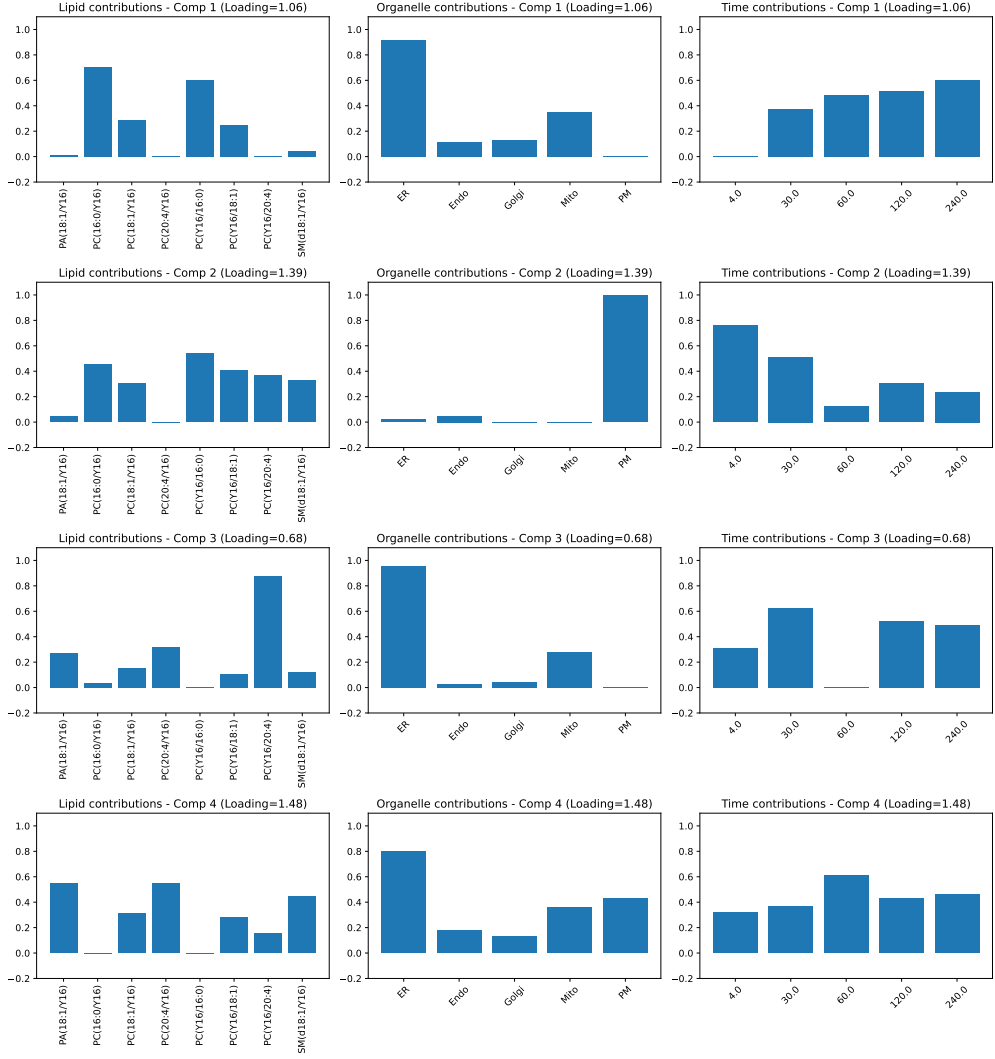}
     \caption{Histograms of the entries of the factors of rank-$4$ CP decomposition with missing values.}
\end{figure}

\section*{Acknowledgments}

The authors thank Neriman Tokcan for useful discussions. The authors thank the MPI-CBG and CSBD for the excellent working conditions. This work started during an internship of IP at the MPI-CBG in Fall 2025. VG is member of the italian group GNSAGA-INdAM. HAH gratefully acknowledges funding from EPSRC EP/Y028872/1 and EP/Z531224/1.


\end{document}